\documentclass[range]{ar2e}

\usepackage{amsmath}
\usepackage{amssymb}
\usepackage{bm}
\usepackage{url}

\usepackage{graphicx}

\begin{document}

\input epsf.tex    

\input psfig.sty


\title{Topological Crystalline Insulators and Topological Superconductors: From Concepts to Materials}

\markboth{Ando \& Fu}{Topological Crystalline Insulators and Topological Superconductors}

\author{Yoichi Ando$^1$ and Liang Fu$^2$
\affiliation{$^1$Institute of Scientific and Industrial Research,
Osaka University, Ibaraki, Osaka 567-0047, Japan\\
$^2$Department of Physics, Massachusetts Institute of 
Technology, Cambridge, MA 02139, USA}}


\begin{abstract}

In this review, we discuss recent progress in the explorations of
topological materials beyond topological insulators; specifically, we
focus on topological crystalline insulators and bulk topological
superconductors. The basic concepts, model Hamiltonians, and novel
electronic properties of these new topological materials are explained.
The key role of symmetries that underlie their topological properties is
elucidated. Key issues in their materials realizations are also
discussed.

\end{abstract}

\maketitle

\section{Introduction}

In the past decade, there has been remarkable progress in our
understanding of topological states of matter. A quantum state may be
called topological when its wavefunctions bear a distinct character that
can be specified by some topological invariant---a discrete quantity
that remains unchanged upon adiabatic deformations of the system.
Materials realizing such topological states in their bulk may be called
topological materials. Since the 1980s, quantum Hall systems \cite{TKNN} and
superfluid Helium 3 (He-3) \cite{Volovik} have been recognized to be
topological, but it was long believed that such topological states are
rather exceptional in nature and exist only in quantum liquids under
extreme conditions (under high magnetic fields or at low temperatures).
However, after the discovery of topological insulators (TIs)
\cite{80, BHZ, Molenkamp, 81, 82, 83, Fu-Kane2007PRB, Hsieh2008, Hasan-Kane, Qi-Zhang, AndoReview}, it has come to be widely
recognized that topological states of matter can actually be widespread.
In this sense, TIs have established a new paradigm about topological
materials. It is generally expected that studies of topological
materials would deepen our understanding of quantum mechanics in solids
in a fundamental way.

By now, the theoretical aspect of TIs are reasonably well understood;
hence, major challenges on the theoretical front concerns expansions of
our notion of topological materials. On the experimental front, however, 
TIs are still far from being sufficiently investigated;
materials issues in three-dimensional (3D) TIs are being solved
\cite{AndoReview}, and there are serious on-going efforts to realize
theoretically predicted novel phenomena, such as topological
magnetoelectric effects \cite{QiTEME} and proximity-induced topological
superconductivity hosting the non-Abelian Majorana zero mode in the vortex
core \cite{FuKaneMajorana}. At the same time, new materials discoveries
are still crucially important in this rapidly developing field, and the
search for new types of topological materials is in progress worldwide
\cite{AndoReview}.

In this review, we mainly focus on two new classes of topological
phases of matter beyond TIs, namely, topological crystalline insulators
(TCIs) and topological superconductors (TSCs). The basic concepts and
effective models are concisely summarized, so that experimentalists can
grasp the essential physics of these topological matters to accomplish 
new material discoveries. We also discuss actual (candidate) materials for
each category and mention the issues to be addressed in future studies.

\subsection{$Z_2$ Topological Insulator}

Before going into the main topics of this review, TCIs and TSCs, let us
briefly summarize the current status of the TI research. As the first
class of materials identified to exhibit topological properties
preserving time-reversal symmetry, TIs are characterized by the
topological invariant called the $Z_2$ index \cite{80, 84}. The concept and definition
of the $Z_2$ index are pedagogically summarized in a recent review
article \cite{AndoReview}. Lately, theoretical works on TIs are mostly
focused on either the effects of electron interactions or the
phenomenological consequences of its topological character. On the
materials front, all TI materials confirmed to date are narrow-gap
semiconductors with an inverted band gap \cite{AndoReview}. Such band
inversions must occur at an odd number of time-reversal-invariant
momenta (TRIMs) to obtain a nontrivial $Z_2$ index. Both two-dimensional
(2D) and 3D TI materials have been studied, and the materials efforts
can be summarized as follows.

Among 2D systems, there are two cases that have been confirmed to be 2D
TIs. Those two 2D TI systems are both realized in artificial quantum
well structure and the band inversion takes place at the center of the
Brillouin zone, i.e. at the $\Gamma$ point. The first system confirmed
to be a 2D TI was a thin layer of HgTe sandwiched by CdTe \cite{BHZ,
Molenkamp}, and the second one was InAs/GaSb heterojunction sandwiched
by AlSb \cite{Zhang_InAsGaSb, Du}. In HgTe, the band inversion is
naturally realized due to the large spin-orbit coupling stemming from
the heavy element Hg, but the cubic symmetry of this material causes the
valence and conduction bands to be degenerate at the $\Gamma$ point; the
quantum confinement and the resulting formation of subbands removes this
degeneracy and leads to the realization of a true 2D TI state. In
InAs/GaSb heterojunction, however, the band inversion is achieved by the
broken (type-III) gap alignment between InAs and GaSb, and the band gap
in the hybridized band structure is created by the anticrossing of the
inverted electron and hole subbands stemming from InAs and GaSb,
respectively. Experimentally, although the materials are difficult to
grow in both cases, it is relatively easy to make the transport through
the helical edge states to become predominant in those 2D TIs by using
electrostatic gating.

Among 3D materials, the first one to be confirmed as 3D TI was
Bi$_{1-x}$Sb$_x$ alloy \cite{Fu-Kane2007PRB, Hsieh2008}. In this
material, the band inversion occurs at three TRIMs, the $L$ points; the
surface band structure is rather complicated, consisting of an odd
number of Dirac cones and additional states \cite{Hsieh2008, Nishide}.
Other 3D TI materials found to date have the band inversion only at the
$\Gamma$ point, and hence they are associated with simpler surface band
structures consisting of a single Dirac cone. Among such simpler 3D TI
materials, the most widely studied are the binary tetradymite compounds
Bi$_2$Se$_3$ and Bi$_2$Te$_3$ \cite{ZhangNP, Chen2009, Xia2009}, in
which the band inversion is due to a strong spin-orbit coupling that
switches the order of two $p_z$-orbital-dominated bands with opposite
parities at the $\Gamma$ point.

Unfortunately, most of the known 3D TI materials are not really
insulating in the bulk due to unintentional doping. Hence, an important
experimental issue has been to find suitable materials that present
sufficiently high bulk resistivity so that the surface transport
properties can be reliably probed \cite{AndoReview}. In this regard, the
ternary tetradymite compound Bi$_2$Te$_2$Se was the first material that
achieved a reasonably large bulk resistivity with a high surface
mobility, allowing clear observations of surface quantum oscillations
\cite{RenBTS, Xiong}. Later, in an alloyed tetradymite compound
Bi$_{2-x}$Sb$_x$Te$_{3-y}$Se$_y$, a series of special compositions to
achieve minimal bulk conduction was identified \cite{RenBSTS}, and the
surface-dominated transport was demonstrated for the first time in bulk
single crystals of a 3D TI \cite{Taskin2010}. In thin films of 3D TIs,
similar surface-dominated transport has been achieved in strained HgTe
\cite{Molenkamp3D} and in Bi$_{2-x}$Sb$_x$Te$_3$ \cite{XueBST}. Also, in
exfoliated thin flakes of Bi$_2$Se$_3$, it was reported that the
deposition of strongly electron-affine molecules called F4TCNQ makes it
possible to achieve the surface-dominated transport \cite{Kim-Fuhrer}.
With those advancements in materials, experimental studies of the
intrinsic properties of 3D TIs have become possible.

\subsection{Symmetry-Protected Topological Phases}

The advent of TIs draws wide attention to the broad notion of
symmetry-protected topological (SPT) phases \cite{85}, of which TI is an example.
Generally speaking, a SPT phase exhibits topological characteristics
(e.g., topological invariants and gapless boundary states) that rely
crucially on the presence of certain symmetry (e.g., time reversal
symmetry), and it can be adiabatically deformed to a trivial phase
(e.g., an atomic insulator) after this underlying symmetry is removed.
In recent years, the search for other SPT phases has attracted
tremendous activities on both theoretical and experimental sides \cite{125}. The
two new topological phases treated in this review, i.e. TCIs and TSCs,
are subsets of SPT phases, and for both of them several material
realizations/candidates are currently under study. Similar to TIs, TCIs
and TSCs are defined by topological invariants encoded in the
wavefunctions of Bloch electrons and of Bogoliubov quasiparticles,
respectively.

\section{Topological Crystalline Insulator}

\subsection{General Concept}

TCIs (TCIs) \cite{Fu} are topological phases of matter that are protected by
crystal symmetries, including rotation, reflection, etc.; for example,
$C_{3v}$ point group requires threefold rotation and reflection
symmetries. A TCI cannot be adiabatically deformed to an atomic
insulator while preserving certain crystal symmetry. Several theoretical
examples of such crystal-symmetry-protected topological phases have been
studied in the context of TIs \cite{80, 81, 82, Teo} and related systems \cite{Mong}. A
systematic search for TCIs requires the classification of topologically
distinct band structures within each crystal class. Given the richness
and complexity of crystallography, the full classification of TCI has
not yet been attained and is an active area of current research. In this
review, we largely focus on a class of TCIs that has been experimentally
realized \cite{HsiehTCI2012, TanakaNP, DziawaNM, XuNC}, whose
topological character is protected by reflection symmetry with respect
to a crystal plane, or equivalently, mirror symmetry. 

Reflection $M$ is equal to a product of spatial inversion $P$ and the
two-fold rotation $C_2$ around the axis perpendicular to the plane of
reflection (hereafter denoted by $z=0$): $M= P C_2 $. In
spin-orbit-coupled systems, $C_2$ is a combined rotation of an electron's
spatial coordinates and spin. 
Thus reflection acts on a spinful wavefunction as follows 
\begin{eqnarray}
M \left( \begin{array}{c}
\psi_\uparrow(\bf r) \\
\psi_\downarrow(\bf r)
\end{array}
\right) = \left( \begin{array}{c}
 - i \psi_\uparrow(\bf \bar r) \\
i \psi_\downarrow(\bf \bar r)
\end{array}
\right), \label{m}
\end{eqnarray}  
where ${\bf \bar r} = (x, y, -z)$. Note that to define $M$ properly
requires picking an orientation for the plane of reflection that
distinguishes the $+z$ and $-z$ directions. 
Owing to the sign reversal of spinor under $2\pi$ rotation, $M^2=-1$ and
hence eigenvalues of $M$ are either $i$ or $-i$. 
 
The presence of mirror symmetry in a crystal has implications for its
energy eigenstates in momentum space, i.e., Bloch states $|\psi_{{\bf
k}} \rangle$. For a 2D crystal that is invariant under $z \rightarrow
-z$, $|\psi_{{\bf k}} \rangle$ can be chosen to be eigenstates of $M$
for all $\bf k$. This yields two classes of Bloch eigenstates with
mirror eigenvalues $\eta=\pm i$, denoted by $|\psi_{{\bf k }, \eta}
\rangle$. For each class of Bloch eigenstates, one can define a
corresponding Chern numbers $N_\eta$. This leads to two independent
topological invariants: the total Chern number $N \equiv N_{+i} +
N_{-i}$ determines the quantized Hall conductance, and a new invariant
called ``mirror Chern number'': $N_M \equiv (N_{+i} - N_{-i})/2$
\cite{Teo}. Importantly, even when the total Chern number is zero, the
mirror Chern number can be a nonzero integer, which then defines a TCI
phase protected by the mirror symmetry. 

The above idea can be generalized to 3D crystals that
have one or multiple mirror planes. The presence of a given mirror
symmetry, say $x \rightarrow -x$, implies that the Bloch states
$|\psi_{{\bf k}} \rangle$ at the $k_x=0$ and $k_x=\pi/a$ planes in the
3D Brillouin zone are mirror eigenstates. Each such mirror-invariant
plane in momentum space is then indexed by its own mirror Chern number.
The complete set of mirror Chern numbers classifies 3D TCI phases with
mirror symmetry. 

The topological character of a TCI leads to gapless states on the
boundary. Importantly, because the boundary can have lower symmetry than
the bulk, not all crystal surfaces of the above 3D TCI are gapless; only
those surfaces that preserve the underlying mirror symmetry are. The
dependence of boundary states on surface orientations is a generic
property of TCIs \cite{Fu, Mong} and enriches topological phenomena in
solids, as we will show below.

\subsection{Models and Materials} 

In 2012, Hsieh {\it et al.} \cite{HsiehTCI2012} predicted the first
class of TCI materials in IV-VI semiconductors, with SnTe as a
representative. These materials crystallize in rock-salt structure. The
symmetry responsible for their topological character comes from the
reflection symmetry with respect to the (110) mirror planes. In stark
contrast, the isostructural compound PbTe in the same IV-VI material
family is predicted to be non-topological. We describe below the
important difference between SnTe and PbTe in electronic structures, and
its implication for TCI. 

Both SnTe and PbTe have small direct band gaps located at four
symmetry-related TRIMs, the $L$ points. The low-energy band structure,
consisting of the doubly degenerate conduction and valence bands in the
vicinity of $L$, is described by a four-band $k\cdot p$ Hamiltonian
$H(\bf k)$ \cite{HsiehTCI2012}, which can be regarded as the low-energy
limit of a microscopic six-band model in the early work of Mitchell \&
Wallis \cite{Wallis}. Alternatively, in the spirit of modern condensed
matter physics, $H(\bf k)$ can be regarded as an effective Hamiltonian,
whose analytical form can be derived entirely from the symmetry
properties of energy bands. The little group that keeps each $L$ point
invariant is $D_{3d}$, a subgroup of the $O_h$ point group of the
rock-salt structure. The group $D_{3d}$ consists of three independent
symmetry operations: spatial inversion ($P$), reflection with respect to
the (110) plane ($M$), and three-fold rotation around the $(111)$ axis
($C_3$). The conduction and valence bands at a given $L$ point form two
sets of Kramers doublets with opposite parity eigenvalues, denoted by
$|\psi^+_{L, \alpha} \rangle$ and $|\psi^-_{L, \alpha}\rangle$
respectively. The two members of a Kramers doublet denoted by $\alpha=1,
2$ have opposite total angular momenta $J_z = \pm \frac{\hbar}{2}$
respectively. Because the axis of rotation is parallel to the plane of
reflection, $J_z$ changes sign under reflection, i.e., $M
|\psi^{\pm}_{L, 1} \rangle = i |\psi^{\pm}_{L, 2} \rangle$ and $M
|\psi^{\pm}_{L, 2} \rangle = i |\psi^{\pm}_{L, 1}\rangle$. 

The above band symmetries dictate the form of the $k \cdot p$
Hamiltonian $H(\bf k)$, where $\bf k$ is measured from a given $L$
point. $H(\bf k)$ is a $4\times 4$ matrix in the basis set of $\{
|\psi^+_{L, 1} \rangle, |\psi^+_{L, 2} \rangle, |\psi^-_{L, 1} \rangle,
|\psi^-_{L, 2} \rangle \}$, which is given by
\begin{eqnarray}
 H({\bf k}) &=& \left( 
 \begin{array}{cccc}
 m & 0 &  -i v' k_z &  -v ( i k_x  + k_y) \\
 0 & m &  v (i k_x -  k_y)  &  - i v' k_z \\
 i v' k_z & -v ( i k_x  + k_y) & -m & 0 \\
v ( i k_x -  k_y)  &  i v' k_z & 0 & -m
 \end{array} \label{hk}
 \right). 
\end{eqnarray}  
$H(\bf k)$ includes all possible terms up to first order in $\bf k$,
which are invariant under the symmetry group $D_{3d}$. Remarkably, after
a rescaling of the coordinate $k_z \rightarrow \frac{v'}{v} k_z$, $H(\bf
k)$ has the same form as the Dirac Hamiltonian in quantum
electrodynamics: $H({\bf k}) = m \Gamma_0 + v \sum_i k_i \Gamma_i$,
where $\Gamma_{0,...,3}$ are Dirac gamma matrices defined by $\Gamma_0=
\sigma_z \otimes I, \Gamma_1= \sigma_x s_y, \Gamma_2 = - \sigma_x s_x$,
and $\Gamma_3= \sigma_y \otimes I$. Thus, the low-energy electronic
properties of SnTe and PbTe are governed by massive Dirac fermions in
3+1 dimension. The conduction and valence bands are separated by an
energy gap $E_g= 2|m|$, and have particle-hole symmetric dispersions
$E_{c,v}(k)= \pm \sqrt{m^2 + v^2 k^2}$.

The topological distinction between SnTe and PbTe arises from their
different Dirac masses. It has long been known that in going from PbTe
to SnTe, the band gap of the alloy Pb$_{1-x}$Sn$_x$Te closes at a
critical Sn composition, $x\sim 0.35$, and then reopens \cite{91}. This band
inversion corresponds to a sign change of the Dirac mass in the
low-energy theory (\ref{hk}). The key insight that led to the prediction
of the TCI phase \cite{HsiehTCI2012} came from the recognition that this
Dirac mass reversal has an important consequence for topology: It
changes the mirror Chern number $N_M$ associated with the $k_x=0$ plane
passing through $\Gamma$ and two $L$ points, such as $\Gamma L_1 L_2$,
$\Gamma L_3 L_4$ and $\Gamma L_1 L_3$ [see Fig. 1(a)]. Energy bands on
these planes are mirror eigenstates indexed by $\eta=-i s_x$. The
simultaneous band inversions at the two $L$ points on the $k_x=0$ plane
add up to an integer value of the mirror Chern number: $1+1=2$.
Therefore, one of the two materials, SnTe or PbTe, must have a nonzero
mirror Chern number $|N_M| = 2$ and thus realizes a TCI phase protected
by mirror symmetry. However, neither material is a $Z_2$ TI
\cite{Fu-Kane2007PRB}, because an even number of band inversions
``annihilate'' each other, as can be seen from addition rule of the
$Z_2$ group classification of time-reversal-invariant systems: $1+1 =0$
mod 2. 

To further determine whether SnTe or PbTe is topologically nontrivial
requires looking into the microscopic band structures of SnTe and PbTe,
which is beyond the scope of the effective theory. Ab initio
calculations show that the conduction (valence) band of PbTe
predominantly comes from the cation Pb (the anion Te) orbitals, as
expected for an ionic insulator made of Pb$^{2+}$ and Te$^{2-}$ in the
atomic limit. In contrast, SnTe displays an anomalous band character: In
a small region of the Brillouin zone around $L$ points, the conduction
(valence) band comes from the anion Te (the cation Sn) orbitals. This
inverted band ordering of SnTe, distinct from an ionic insulator, is
responsible for the experimentally observed decrease (increase) of band
gap under tensile strain or pressure, which increases (decreases) the
lattice constant towards (away from) the atomic limit. Putting together
the results of the low-energy theory, topological band theory, and ab
initio calculation, Hsieh et al. predicted that SnTe is a TCI, whereas
PbTe is not.

\subsection{Topological Crystalline Insulator Surface States}

The nonzero mirror Chern number in the SnTe class of TCIs guarantees the
existence of topological surface states on crystal faces that are
symmetric with respect to the (110) mirror planes. Such crystal faces
have a Miller index ($hhk$). (The cubic symmetry of SnTe dictates that the
situation is the same for ($khh$) and ($hkh$) faces.) Three common surface
terminations of IV-VI semiconductors are (001), (111), and (110) [see
Fig. 1(a)], which all satisfy this condition. Interestingly, depending
on the surface orientation, there are two types of TCI surface states,
with qualitatively different electronic properties, as schematically
shown in Fig. 1(b). 

The first type of TCI surface states exist on
the (001) and (110) surface, where a pair of $L$ points are projected
onto the same TRIMs on the surface. For the (001) surface, $L_1$ and
$L_2$ are projected onto $X_1$, and $L_3$ and $L_4$ are projected onto
$X_2$. In this case, the two massless surface Dirac fermions resulting
from band inversions at $L_1$ and $L_2$ ($L_3$ and $L_4$) hybridize with
each other at the surface and create unprecedented surface states at
$\bar{X_1}$ ($\bar{X_2}$) with a double-Dirac-cone band structure. The
essential properties of these surface states are captured by the
following minimal $k \cdot p$ model at a given $\bar{X}$ point
\cite{Liu-Duan-Fu}: 
\begin{eqnarray}
H_{\bar{X}}({\bf k}) =  (v_x k_x   s_y - v_y k_y s_x ) \otimes I 
 + m  \tau_x + \delta s_x  \tau_y. \label{001}
\end{eqnarray}    
Here, the first term describes two identical copies of surface Dirac
fermions associated with $L_1$ and $L_2$ (denoted by $\tau_z= \pm 1$),
respectively; the other terms describe all possible inter-valley
hybridizations to zeroth order in $\bf k$, which satisfy all the
symmetries of the (001) surface \cite{Fang, Hsin, 134}. The calculated
surface band structure of $H_{\bar{X}}$, plotted in Fig. 1(c), shows
many interesting features. At low energy close to the middle of the bulk
gap, the surface states consist of a pair of Dirac cones located
symmetrically away from $\bar{X}$ on the line $\bar{X} \bar{\Gamma} $.
The corresponding Fermi surface is two disconnected elliptical Dirac
pockets. As the Fermi energy increases, these two pockets become
crescent-shaped, touch each other on the line $\bar{X} \bar{M}$, and
reconnect to form a large electron pocket and a small hole pocket, both
centered at $\bar{X}$. This change of Fermi surface topology from being
disconnected to connected, known as Lifshitz transition, leads to a
Van-Hove singularity in the density of states at the transition point. 

The surface band structures discussed above are directly related to the
mirror Chern number of TCIs. The (001) surface exhibits surface band
crossings on the line $\bar{X} \bar{\Gamma}$ between bands of opposite
mirror eigenvalues, and the (111) surface shows similar crossings on the
line $\bar{\Gamma} \bar{M}$. These crossings protected by mirror
symmetry guarantee the gapless nature of TCI surface states, replacing
the role of Kramers degeneracy in $Z_2$ TIs. The fact that two surface
band crossings can take place at any point on the entire line agrees
with the mirror Chern number $|N_M|=2$, which precisely illustrates the
principle of bulk-edge correspondence in topological phases of matter. 

The second type of surface states exists on the (111) surface. Here one
of the four $L$ points in the bulk projects to the $\bar{\Gamma}$ point
on the surface Brillouin zone, and the other three $L$ points project to
$\bar{M}$. As expected from the effective theory of band inversion, the
(111) surface consists of four branches of massless Dirac fermions: one
branch located at $\bar{\Gamma}$ and three at $\bar{M}$. Importantly,
the mirror symmetry guarantees that such surface states are connected in
a topologically nontrivial manner along the mirror-invariant line
$\bar{\Gamma}-\bar{M}$, such that they cannot be removed. Similar to the
free (111) surface, symmetry-protected interface states should exist on
the (111) heterostructure between SnTe and PbTe. These interface states
were anticipated from early field-theoretic studies \cite{92, 93}. The
discovery of TCIs has now revealed that these states (so far unobserved)
stem from the TCI material SnTe (but not PbTe) and are topologically
equivalent to its (111) surface states. 

\subsection{Experiments}

Following the prediction by Hsieh {\it et al.} \cite{HsiehTCI2012} 
that SnTe is a TCI, angle-resolved photoemission
spectroscopy (ARPES) experiments showed that SnTe \cite{TanakaNP} and
Pb$_{1-x}$Sn$_x$Se \cite{DziawaNM} are indeed a new type of topological
materials characterized by peculiar surface states consisting of {\it
four} Dirac cones. Later, the TCI phase was confirmed to remain in the
Pb$_{1-x}$Sn$_x$Te alloy for $x \gtrsim$ 0.25 \cite{XuNC, TanakaPbSnTe}.
Those materials crystallize in the cubic rock-salt structure, which can
be cleaved along either (001) or (111) planes. Like in $Z_2$ TIs, the
four Dirac cones of TCIs are spin non-degenerate and are helically
spin-polarized \cite{XuNC}.

The initial experiments done on the (001) surface \cite{TanakaNP,
DziawaNM, XuNC, TanakaPbSnTe} found a double-Dirac-cone structure near
the $\bar{X}$ point of the surface Brillouin zone [Fig. 2].
Remarkably, the Dirac points of the surface state are {\it not} located
at the TRIMs; rather, their locations are restricted on the mirror axes
of the surface Brillouin zone. Such a surface state structure stems from
a mirror-symmetry-constrained hybridization of two Dirac cones
\cite{HsiehTCI2012}, as described in Section 2.3. The predicted Lifshitz
transition stemming from the merger of two nearby Dirac cones has also
been experimentally observed \cite{TanakaNP, DziawaNM, XuNC}.

On the (111) surface, however, all four Dirac points are located at the
TRIMs: one at $\bar{\Gamma}$ and three at $\bar{M}$ points [Fig. 1(b)].
It was found both by {\it ab initio} calculations \cite{Liu-Duan-Fu,
Safaei(111)} and by experiments \cite{Tanaka(111)} that the energy
location of the Dirac point as well as the Fermi velocity are different
for the Dirac cones at $\bar{\Gamma}$ and $\bar{M}$. These two kinds of
Dirac cones were found to manifest themselves in the surface transport
properties as different components of the surface quantum oscillations
observed in SnTe thin films grown along the [111] direction
\cite{TaskinSnTe}.

Interestingly, the difference in the Dirac cones at $\bar{\Gamma}$ and
$\bar{M}$ introduces peculiar valley degrees of freedom and makes it
possible to conceive unique valleytronics for the (111) surface states
of TCIs \cite{EzawaTCI}. Another type of valley-dependent phenomenon
arises on the (001) surface due to a spontaneous structural distortion
that selectively breaks mirror symmetries and opens gaps at the four
Dirac valleys \cite{HsiehTCI2012}. Indeed, such gap openings at two of the four Dirac
cones have been observed in the Landau level map of Pb$_{1-x}$Sn$_x$Se
(001) surface states using scanning tunneling microscope (STM)
\cite{OkadaScience}.

\subsection{Perturbations to the Topological Crystalline Insulator Surface States}

Compared with TIs, TCI surface states have a much wider range
of {\it tunable} electronic properties under various perturbations, such
as structural distortion, magnetic dopant, mechanical strain, thickness
engineering, and disorder \cite{HsiehTCI2012, TanakaPbSnTe, SerbynFu, TangFu,
LiuDevice, QianJuFu, FuKane2012}. We now briefly discuss
the interesting consequences of these perturbations on the (001) surface
states, some of which have been experimentally observed. 

(i) {\it Ferroelectric structural distortion}: A common type of
structural distortion in IV-VI semiconductors is a relative displacement
$\bf u$ of the cation and anion sublattices [Fig. 3(a)], which leads to
a net ferroelectric polarization. Depending on the direction of $\bf u$,
this distortion breaks the mirror symmetry with respect to either one or
two mirror planes, and therefore generates nonzero mass for the original
massless Dirac fermions on the TCI (001) surface. Both the magnitudes
and the {\it signs} of the Dirac masses at the four valleys depend on
the direction of $\bf u$, resulting in a rich phase diagram [Fig. 3(a)]
\cite{HsiehTCI2012}. This Dirac mass generation by symmetry breaking has
been observed in a scanning tunneling microscopy experiment on the TCI
Pb$_{1-x}$Sn$_x$Se \cite{OkadaScience} as already mentioned. 

(ii) {\it Magnetic dopant}: The exchange coupling of TCI surface Dirac
fermions and magnetic moments of dopants results in
time-reversal-symmetry breaking. In particular, an out-of-plane
magnetization opens up Zeeman gaps of the same signs at the four Dirac
points, leading to quantum Hall effect with $\sigma_{xy}= 4 \times
\frac{e^2}{2h} $ \cite{HsiehTCI2012}. This offers a promising route to
quantum anomalous Hall states in TCI thin films, with large quantized
Hall conductance \cite{FangBernevig}. 

(iii) {\it Mechanical strain}: The Dirac points on the (001) surface of
TCIs are not pinned to TRIMs as in the case of TIs. In this case, a
mechanical strain can shift the Dirac point positions in $\bf k$ space
in a similar way as an electromagnetic gauge field acts on an
electron \cite{SerbynFu}. As a result, a nonuniform strain field
generates a nonzero pseudo-magnetic field, that can dramatically alter
the electronic properties of TCI surface states. It has been
proposed \cite{TangFu} that a Landau-level-like flat band can be created
by a periodic strain field due to the dislocation array that
spontaneously forms on the interface of TCI heterostructures, and the
resulting high density of states may be responsible for the unusual
interface superconductivity found in these systems \cite{interfaceSC}. 

(iv) {\it Thickness engineering}: In (001) thin films
of TCIs, the top and bottom surface states hybridize to open up an
energy gap at the Dirac points. However, the inverted band structure at
the $X$ points remains down to a few layers. In the wide range of
intermediate thickness, these films realize a two-dimensional TCI phase
that has spin-filtered edge states \cite{LiuDevice}. Unlike quantum spin Hall insulators,
these edge modes consist of an even number of Kramers pairs, which are
protected by the symmetry with respect to the film's middle plane.
Applying a small out-of-plane electric field breaks this mirror symmetry
and hence gaps out these spin-filtered edge states \cite{LiuDevice}. This electrically
tunable edge channel may be regarded as a topological transistor, whose
ON and OFF states are controlled by an electrically induced gap in the
topological edge channel, instead of carrier injection/depletion (see
Fig. 4). 

(v) {\it Disorder}: Unlike internal symmetries (such as time-reversal
symmetry), spatial symmetries (such as mirror symmetry) are always
violated in the presence of disorder \cite{86}. This raises the question of
whether TCI phases are stable against disorder. It has been argued that
the topological surface states in the SnTe class of TCIs cannot be
localized even under strong disorder, provided that time reversal
symmetry is present \cite{HsiehTCI2012, FuKane2012}. This remarkable
absence of localization is protected by the restored mirror symmetry
after disorder averaging, or average mirror symmetry. Intuitively, one
can treat the strongly disordered TCI surface as an ensemble of domains,
where each domain breaks mirror symmetry and hence is locally gapped.
However, there exist two types of domains that are related to each other
by mirror symmetry. As a unique property of TCIs, the interface between
the two mirror-related domains hosts a single one-dimensional helical
mode \cite{HsiehTCI2012}. Since time-reversal-symmetry forbids backscattering within
helical states, each domain wall is a ballistic conductor. 
The average mirror symmetry further guarantees that the two types of
domains occur with equal probability. As a result, the conducting domain
wall percolates throughout the entire surface, leading to
delocalization. To make the above argument rigorous, we now present a
proof \cite{94} that in the presence of time reversal symmetry, the
disordered TCI surface cannot be localized. Let us consider a setup
shown in Fig. 3(b), where a disordered region of the TCI surface is
confined between two gapped regions on its left and right, which are
obtained by externally breaking the mirror symmetry and are swapped
under mirror operation. In this setup, the disordered region is
topologically equivalent to a domain wall between the two gapped
regions, and hence hosts a single delocalized one-dimensional helical
mode. Now suppose this disordered region could be localized, this
helical mode, which cannot be ``split", must sit either on the left or
right boundary, which contradicts with the mirror symmetry of the entire
setup. This proves by contradiction that the disordered TCI surface
cannot be localized.

It is now understood that the delocalization of boundary states due to
protection from an average symmetry occurs in a much broader class of
topological phases, which include for example weak topological
insulators that are protected by translational symmetry \cite{127, 128, 129, 130}. Like
the TCI with mirror symmetry, these topological phases (termed
``statistical topological insulators" \cite{130}) lie beyond the tenfold
classification scheme \cite{Ludwig}, and have the common defining property that
their boundary states exhibit two topologically distinct phases when the
underlying symmetry is explicitly broken in opposite ways. It then
follows from the argument presented above that when the symmetry is
preserved on average, the disordered surface precisely sits at a
topology-changing phase transition point and for this reason cannot be
localized. The physics of such delocalization due to topology and
average symmetry has been precisely formulated in a field-theoretic
approach to Anderson localization \cite{FuKane2012}, and confirmed in numerical
studies \cite{128, 131, 132, 133}.

\section{Topological Superconductor}

\subsection{General Concept}

TSCs can be regarded as a superconducting cousin of TIs. Unlike
insulators whose total number of electrons is conserved, superconductors
(and superfluids) spontaneously break the $U(1)$ symmetry associated
with the fermion number conservation. Instead, only the fermion number
parity (i.e., even or odd) is conserved in the mean-field theory of
superconductivity. This important difference in symmetry called for a
new topological classification of superconductors different from
insulators, which was systematically obtained in Refs. \cite{Ludwig,
Kitaev} and led to the theoretical finding of a wide class of TSCs. 
Several concrete examples of TSC appeared in early models
studies by Read \& Green \cite{ReadGreen} and Kitaev \cite{Kitaevchain},
as well as others \cite{Raghu, Roy}. The search for TSCs in real
materials is currently an exciting research endeavor in condensed matter
physics.

A TSC is most easily conceived in a fully-gapped superconductor as one
that cannot be adiabatically connected to a Bose-Einstein condensate
(BEC) of Cooper pairs, in the same sense that a TI cannot be
adiabatically deformed to the atomic limit. By this standard,
conventional $s$-wave spin-singlet superconductors are clearly
non-topological, because they exhibit a smooth crossover from the
weak-coupling Bardeen-Cooper-Schrieffer (BCS) limit to the
strong-coupling BEC limit without undergoing a gap-closing phase
transition. This implies that unconventional pairing symmetry is a
necessary (but not sufficient) condition for TSCs. Although the concept
of TSC is most transparent in fully gapped superconductors, it is
important to note that nodal (zero-gap) superconductors can also be
topological as long as a topological invariant is well defined; indeed,
for several particular cases of nodal superconductors, topological
classifications have been accomplished and concrete topological
invariants are found \cite{Sasaki2011, Sato_nodal}.

As a consequence of its nontrivial topology, irrespective of whether it is 
fully gapped or nodal, a TSC is guaranteed to possess protected gapless
excitations on the boundary. Importantly, unlike in TIs, these
excitations are not electrons or holes (as in a normal metal) but
Bogoliubov quasiparticles, namely, coherent superpositions of electrons
and holes. The corresponding surface states are Andreev bound states.

The classification of TSCs and the nature of their surface Andreev bound
states depend crucially on the presence or absence of internal
symmetries such as time reversal and spin rotation. Of particular
interests are time-reversal-breaking TSCs (the superconducting cousin of
quantum Hall insulators) and time-reversal-invariant TSCs [the
superconducting analog of TIs, \cite{87}]. A famous example of the former type is a
2D chiral $p_x + i p_y$ spin-triplet superconductor. There is evidence
that an extensively studied material, Sr$_2$RuO$_4$, is a $p$-wave
superconductor, but there is no consensus as to whether it fulfills all the
requirements of a chiral TSC \cite{Maeno}, although there is
experimental indication of surface Andreev bound states
\cite{Kashiwaya}. 

In the following, we focus on time-reversal-invariant TSCs in
spin-orbit-coupled systems, which have attracted wide attention only
recently. Remarkably, the gapless quasiparticles on the surface of such
TSCs do not carry conserved quantum numbers associated with an
electron's charge or spin, and are completely indistinguishable from
their antiparticles. Because particles that are their own antiparticles
are called Majorana fermions \cite{Alicea, Beenakker}, the
quasiparticles on the surface of those TSC are emergent helical Majorana
fermions in the solid state, which can be thought of as one half of the
helical Dirac fermion on a TI surface.

\subsection{Odd-Parity Criterion}

Fully gapped, time-reversal-invariant TSCs are indexed by a $Z_2$
topological invariant in one and two dimensions, and by an integer
invariant in three dimensions \cite{Ludwig, Kitaev}. For a given
superconductor, the value of its topological index can in principle be
calculated from the band structure and pair potential, using explicit
but complicated formulas. Alternatively, when the superconducting energy
gap $\Delta$ is much smaller than the Fermi energy $\mu$ (which holds
for most known superconductors), the topological index is entirely
governed by the topology of the normal-state Fermi surface and the
symmetry of the superconducting order parameter, without reference to
the full band structure in the Brillouin zone. As we show below,
this Fermi surface and pairing-symmetry-based approach provides a
straightforward criterion for TSCs that is conceptually transparent
and experimentally accessible. 

In particular, the criterion for time-reversal-invariant TSCs becomes
remarkably simple for materials with inversion symmetry. In this case,
the pairing order parameter is either even or odd parity. In the absence
of spin-orbit coupling, even-parity pairing corresponds to spin-singlet
pairing, whereas odd-parity pairing corresponds to spin-triplet pairing.
When spin-orbit coupling is present, the notions of spin-singlet and
-triplet pairing are no longer well-defined, but there remains a sharp
distinction between even- and odd-parity pairings. It was found
\cite{Fu-Berg, Sato} that even-parity pairing inevitably leads to
topologically trivial superconductors, whereas odd-parity pairing leads
to TSCs under broad conditions of Fermi surface topology. Specifically,
when the Fermi surface encloses an odd number of TRIMs, odd-parity
pairing is guaranteed to create topological superconductivity. This
criterion for TSCs holds in all three spatial dimensions and is proven
by generalizing the parity criterion for TIs \cite{Fu-Kane2007PRB} to
superconductors. In addition, 3D TSCs can also be realized when the
Fermi surface encloses an even number of TRIMs, provided that the
odd-parity order parameters on different Fermi pockets have the same
sign \cite{Qi-Fu}. 

The intimate connection between odd-parity pairing and topological
superconductivity can be intuitively understood by analyzing the
transition from the weak-coupling BCS regime to the strong-coupling BEC
regime in a simple one-band system as the pairing interaction increases.
Both regimes can be treated by mean-field theory. In the BCS regime, the
chemical potential $\mu$ is inside the energy band and is much larger
than the pairing gap, whereas in the BEC regime, $\mu$ lies below the
band bottom. Therefore, the BCS-BEC transition takes place when the
chemical potential (as determined self-consistently) coincides with the
band edge and the Fermi surface shrinks to a point at ${\bf k}=0$.
Importantly, the odd-parity pair potential, a $2 \times 2$ matrix in the
space of the doubly degenerate energy band, must satisfy
$\Delta(\mathbf{k}) = -\Delta (-\mathbf{k})$ and hence is guaranteed to
vanish at ${\bf k}=0$. As a result, right at this BCS-BEC transition
point, the quasiparticle dispersion becomes gapless at ${\bf k}=0$. This
unavoidable gap closing implies that the BCS regime of odd-parity
superconductors cannot be adiabatically connected to the topological
trivial BEC regime and hence must be topologically nontrivial. In
contrast, for even-parity superconductors, the pairing gap
$\Delta(\mathbf{k}) = \Delta (-\mathbf{k})$ can stay finite at ${\bf
k}=0$, and therefore the BCS regime is adiabatically connected to the
BEC regime and hence is topologically trivial. This argument clearly
demonstrates that odd-parity pairing is the key requirement of TSCs.

\subsection{Material Proposals}

Although odd-parity pairing has long been known in the context of $p$-wave
superfluid He-3, odd-parity superconductivity is rare in solid-state
systems. Prime examples are Sr$_2$RuO$_4$ \cite{Maeno} and certain heavy
fermion superconductors [e.g., UPt$_3$ \cite{TailleferRMP}], where the
driving force for odd-parity pairing comes from the strong electron
correlation in $d$ or $f$ orbitals. However, these odd-parity
superconductors appear to break time-reversal symmetry and hence do not
qualify as time-reversal-invariant TSCs. 

In searching for time-reversal-invariant TSCs, Fu \& Berg \cite{Fu-Berg}
proposed a new mechanism for odd-parity pairing facilitated by strong
spin-orbit coupling, as well as a possible realization of this mechanism
in a candidate material Cu$_x$Bi$_2$Se$_3$. The main idea is simple:
Strong spin-orbit coupling locks an electron's spin to its momentum and
orbital component and thereby converts a bare interaction that is
short-ranged and spin-independent to an effective interaction between
Bloch electrons that is both spin- and momentum-dependent. With such a
nontrivial spin- and momentum-dependence, this effective interaction is
then capable of generating odd-parity superconductivity. 

Cu$_x$Bi$_2$Se$_3$ is a doped TI that was recently
found to be superconducting, with a maximum transition temperature of
3.8 K \cite{Hor}. The proposed odd-parity pairing in Cu$_x$Bi$_2$Se$_3$
is based on a microscopic two-orbital model of its ferminology, which
also provides a minimal description of spin-orbit coupling in the
presence of inversion symmetry. Unlike the Rashba spin-splitting caused
by inversion asymmetry, spin-orbit coupling in centrosymmetric materials
arises from the interplay between an electron's spin, atomic orbitals, and
crystalline anisotropy, and its form depends on crystal symmetry. The
two relevant orbitals in Cu$_x$Bi$_2$Se$_3$ are located on the upper and
lower part of the quintuple layer, respectively; hence, the electronic
structure can be modeled as a stack of bilayer unit cells along the
$z$ axis. On a given layer, there is a structural asymmetry between $z$
and $-z$, which leads to a local electric field that points
perpendicular to the plane in opposite directions on the top and bottom
layer. This electric field generates a Rashba spin-orbit coupling
associated with electron's motion within each plane, 
\begin{eqnarray}
H_{\rm soc} = v \sigma_z  (k_x s_y - k_y s_x),
\end{eqnarray}
which has opposite signs for the two orbitals, labeled by $\sigma_z =\pm
1$. In addition, inter-plane hopping along the $z$ direction connects
the two orbitals in a staggered way similar to the Su-Heeger-Schrieffer
model for polyacetylene. Taking both intra- and inter-plane motion into
account, we arrive at the following Hamiltonian for the normal state of
Cu$_x$Bi$_2$Se$_3$:
\begin{eqnarray}
H_{\rm 3D}=  v \sigma_z  (k_x s_y - k_y s_x) + v_z k_z \sigma_y + m \sigma_x, \label{h3d}
\end{eqnarray}  
which captures the low-energy band structure near the $\Gamma$ point up
to first order in $\bf k$. It is worth pointing out that apart from a
change of orbital basis, the low-energy Hamiltonian Eq. (\ref{h3d}) for
Cu$_x$Bi$_2$Se$_3$ takes an identical form as the Hamiltonian for SnTe,
Eq. (\ref{hk}), because both are determined by the $D_{3d}$ point group (or
little group) of the crystal. 

Cu$_x$Bi$_2$Se$_3$ appears to be a weakly or moderately correlated
electron system. The parent compound Bi$_2$Se$_3$ is a naturally doped
semiconductor consisting of extended $p$-orbitals, and Cu-doping leads
to a rigid-band shift of the Fermi level deeper into the conduction
band. Fu \& Berg \cite{Fu-Berg} studied superconductivity within the two-orbital model
of Cu$_x$Bi$_2$Se$_3$ expressed in Eq. (\ref{h3d}), assuming that the
pairing interaction is short-ranged in space, as in the standard
treatment of weak-coupling superconductors. Under this assumption, the
pair potential is momentum-independent, but can have a nontrivial
internal structure with spin-orbit entanglement, which is not possible
in a single-orbital system. On the basis of symmetry classification,
four types of such pairings were found and listed in Table 1; each one
has a different symmetry corresponding to the irreducible
representations of the $D_{3d}$ point group $A_{1g}$, $A_{1u}$,
$A_{2u}$, and $E_u$, respectively. The $A_{1g}$ pairing is even parity
and conventional $s$-wave, whereas all others are odd parity and
unconventional. Specifically, the $A_{2u}$ pairing is intra-orbital
spin-singlet, but has opposite signs on the two orbitals. Both $A_{1u}$
and $E_u$ pairings are orbital-singlet and spin-triplet: The former has
zero total spin along the $z$ axis, whereas the latter has zero total
spin along an in-plane direction, spontaneously breaking the three-fold
rotation symmetry of the crystal. Because of the spin-orbit coupling,
these two spin-triplet pairings are non-degenerate. 

In discussing the likely pairing symmetry of Cu$_x$Bi$_2$Se$_3$, Fu \&
Berg \cite{Fu-Berg} studied the phase diagram of the two-orbital model under attractive
density-density interactions, which could come from electron-phonon
coupling. The mean-field calculation showed that the $s$-wave pairing is
favored when the intra-orbital attraction $U$ exceeds the inter-orbital
one $V$, whereas the odd-parity $A_{1u}$ pairing is favored when the
inter-orbital attraction is stronger. It is remarkable that
unconventional odd-parity pairing can be realized in a model with purely
attractive and short-range interactions, which is made possible by the
strong spin-orbit interaction comparable to the Fermi energy, as one can
see in Eq. (\ref{h3d}). The requirement of $V>U$ may be achieved by
taking into account the reduction of phonon-meditated attraction by
renormalized Coulomb repulsion, which is larger for electrons occupying
the same orbital. 

The $A_{1u}$ odd-parity pairing generates a full superconducting energy
gap over the elliptical Fermi surface that encloses the
time-reversal-invariant momentum $\Gamma$, thereby satisfying all the
requirements for 3D time-reversal-invariant TSCs. Indeed, the $A_{1u}$
superconducting phase supports two-dimensional massless helical Majorana
fermions on the surface \cite{Fu-Berg}, which exhibits a novel
energy-momentum dispersion \cite{Hsieh-Fu, Lee, Tanaka}. The other two
odd-parity phases, $A_{2u}$ and $E_u$, have point nodes. Nonetheless,
both phases also have Majorana fermion surface states with Fermi arcs,
whose existence is related to certain weak topological invariants
\cite{Sato_nodal}.

\subsection{Experiments and Open Issues
}

As is discussed above, superconductors derived from
TIs are interesting candidates for bulk TSCs. Among
them, the most widely studied material has been Cu$_x$Bi$_2$Se$_3$,
which actually provided the motivation for the Fu-Berg theory
\cite{Fu-Berg}. The superconductivity in this material shows up as a
result of Cu intercalation to the van der Waals gap of the parent
Bi$_2$Se$_3$ compound \cite{Hor}. The bulk carrier density $n_{\rm 3D}$
of Cu$_x$Bi$_2$Se$_3$ is very low for a superconductor, $n_{\rm 3D}
\simeq$ 1 $\times$ 10$^{20}$ cm$^{-3}$; for such a low carrier density,
the maximum $T_c$ of 3.8 K in Cu$_x$Bi$_2$Se$_3$ is anomalously high
within the context of the BCS theory, in which $T_c$ is exponentially
diminished as the density of states at the Fermi energy is reduced. As a
matter of fact, the BCS theory predicts an order of magnitude lower
$T_c$ for such a low $n_{\rm 3D}$ \cite{Cohen}, and indeed, the
prototypical low-carrier-density superconductor SrTiO$_3$ has the
maximum $T_c$ of 0.5 K for $n_{\rm 3D} \simeq$ 1 $\times$ 10$^{20}$
cm$^{-3}$ \cite{SrTiO3}. The anomalously high $T_c$ for the very low
carrier density is one of the possible indications of an unusual
electron pairing in Cu$_x$Bi$_2$Se$_3$.

Superconducting Cu$_x$Bi$_2$Se$_3$ is difficult to be synthesized with the
usual melt-growth technique \cite{Hor}, but an electrochemical technique
makes it possible to synthesize samples with the superconducting volume
fraction up to $\sim$70\% near $x \simeq$ 0.3 \cite{Kriener2011PRL}.
Using such high-volume-fraction samples, it was found using specific-heat
measurements that this material is likely to have a fully gapped
superconducting state without gap nodes \cite{Kriener2011PRL}. More
importantly, point-contact spectroscopy experiments found signatures of
Andreev bound states [Fig. 5(a)], which points to the realization of
unconventional odd-parity superconductivity, meaning that
Cu$_x$Bi$_2$Se$_3$ is a bulk TSC \cite{Sasaki2011}. Theoretically, the
surface Andreev bound states of such a bulk TSC are nothing but the
helical Majorana fermion state. Therefore, the point-contact experiments
may have seen a signature of Majorana fermions. 

However, an STM study of Cu$_x$Bi$_2$Se$_3$ found only a conventional
tunneling spectrum \cite{Levy}, which created a controversy regarding
the nature of superconductivity in Cu$_x$Bi$_2$Se$_3$. In this context,
it is worth noting that recent self-consistent calculations of the local
density of states (LDOS) in Cu$_x$Bi$_2$Se$_3$ concluded that the
existence of the topological surface state must give rise to a two-gap
structure in the LDOS spectrum at the surface if the bulk
superconducting state is of the conventional BCS type
\cite{Mizushima2014}. Therefore, it is not so straightforward to understand 
the STM result.

Recently, an ARPES study found that the Fermi surface of superconducting
Cu$_x$Bi$_2$Se$_3$ is a warped cylinder and hence the system is
essentially quasi-2D \cite{88}. This result suggests the possibility that this
material is actually a 2D TSC and the topological boundary states exist
only on the side surface. If so, the point-contact spectroscopy using
silver nanoparticles \cite{Sasaki2011} could have probed the Andreev
bound states at the terrace edges, whereas the STM measurements on the
top surface would not probe any boundary states \cite{Mizushima2014}. Clearly, further
studies of Cu$_x$Bi$_2$Se$_3$ using different techniques, such as
nuclear magnetic resonance or $\pi$ junctions, are desirable for
elucidating the true pairing symmetry.

Another interesting candidate of a bulk TSC is superconducting In-doped
SnTe \cite{Sasaki2012}, which is a hole-doped TCI preserving the
topological surface states even after the In doping \cite{SatoInSnTe}.
The effective Hamiltonian of this system has essentially the same form
as that of the 3D version of Cu$_x$Bi$_2$Se$_3$, and hence the symmetry
classification of the possible gap functions in the Fu-Berg theory
\cite{Fu-Berg} still applies. This means that the strong spin-orbit
coupling needed to make SnTe topological may also lead to unconventional
superconductivity in Sn$_{1-x}$In$_x$Te by promoting Cooper pairing
between two different orbitals with opposite parity. Intriguingly, the
point-contact spectroscopy of Sn$_{1-x}$In$_x$Te found signatures of
surface Andreev bound states [Fig. 5(b)] \cite{Sasaki2012} similar to
those found in Cu$_x$Bi$_2$Se$_3$, pointing to the realization of a
topological superconducting state.

It is prudent to mention that the In-doping dependence of $T_c$ in this
material is complicated \cite{Novak2013}, and it has been suggested that
topological superconductivity is realized only in a narrow range of In
content near 4\% where disorder becomes minimal \cite{Novak2013}. The
specific-heat measurements of Sn$_{1-x}$In$_x$Te found that the
superconducting state is fully gapped and the volume fraction is
essentially 100\% \cite{Novak2013}. The absence of impurity phases in
Sn$_{1-x}$In$_x$Te is an advantage, compared with Cu$_x$Bi$_2$Se$_3$, for
elucidating the nature of pairing symmetry. If the bulk is indeed
topological, the surface Andreev bound states of Sn$_{1-x}$In$_x$Te
consist of four valleys of helical Majorana fermions because there are
four bulk Fermi pockets located at the $L$ points.

Very recently, a new TI-based superconductor,
Cu$_x$(PbSe)$_5$(Bi$_2$Se$_3$)$_6$ (hereafter called CPSBS), was
discovered \cite{Sasaki2014}. This material is interesting in that its
specific-heat behavior strongly suggests the existence of gap nodes, and
hence this is almost certainly an unconventional superconductor. The
building block of the crystal structure of (PbSe)$_5$(Bi$_2$Se$_3$)$_6$
consists of two quintuple layers (QLs) of Bi$_2$Se$_3$ separated by 
one-unit-cell-thick PbSe, and hence it can be called a naturally formed
heterostructure of alternating topological and non-topological units; to
make it a superconductor, Cu is intercalated into the van der Waals gap
between the two QLs of the Bi$_2$Se$_3$ units. Although the unconventional
nature in Cu$_x$Bi$_2$Se$_3$ and Sn$_{1-x}$In$_x$Te has so far been
inferred only through the surface properties, the unconventional
superconductivity in CPSBS is indicated by bulk properties [Fig. 5(c)].
This material has a quasi-2D Fermi surface, so the existence of gap
nodes leads to the appearance of surface Andreev bound states on some
particular planes that are parallel to the $c^*$ axis \cite{89}. Importantly, the
strong spin-orbit coupling coming from the Bi$_2$Se$_3$ unit makes the
Andreev bound states spin-split and form a Kramers pair. The resulting
spin-nondegenerate surface states can be identified as helical Majorana
fermion states.

It is noteworthy that any spin-triplet superconductor is potentially a
bulk TSC, either gapped or gapless. Hence, well-established triplet
superconductors such as Sr$_2$RuO$_4$ \cite{Maeno} and UPt$_3$
\cite{TailleferRMP} may well be topological, but their exact topological
nature remains to be identified. Note that a topological bulk state
leads to the appearance of surface Andreev bound states [Fig. 5(d),
\cite{Kashiwaya}] as topological gapless quasiparticle states. However,
in those spin-triplet TSCs, the surface Andreev bound states may not be
identified as Majorana fermion states, because they are spin degenerate;
remember, two Majorana fermions with the same $\mathbf{k}$ can form a
complex linear combination to result in an ordinary fermion.
Nevertheless, a Majorana zero mode is expected to show up in the core of
half-quantized vortices that are peculiar to triplet superconductors
having the $d$-vector degrees of freedom \cite{Maeno}.

\section{Outlook}

The discovery of TIs initiated a new trend to pursue topologically
nontrivial phases in quantum materials, and one would expect this new
trend to keep producing fundamental discoveries about novel quantum
phases of matter characterized by nontrivial topologies. As is
emphasized in the present review, important ingredients for the
theoretical investigations of new types of topological materials are the
construction of effective models and the symmetry analysis of such
models. In this respect, theoretical imaginations to conceive exotic
models are obviously important, but perhaps more important is to
find/design realistic materials to realize such models so that the
theoretical predictions can be verified by experiments. In any case,
because topologies can only be analyzed mathematically in concrete models,
the discoveries of materials characterized by new topologies are
necessarily led by theoretical insights.

On the experimental front, besides exploring new kinds of topological
materials, it is important to establish practical understanding of known
topological materials and to elucidate peculiar phenomena associated
with such materials. In this respect, the implications of the valley
degrees of freedom in TCIs on various physical properties are worth
pursuing. For such efforts, availability of high-quality thin films,
whose Fermi level can be gate controlled, would be crucially important.
Regarding TSCs, the physics of extended and dispersive Majorana fermions
on the surface of certain TSCs is a new area of research and may yield
rich phenomenology, as was the case with massless Dirac fermions in
graphene \cite{90}. Also, finding ways to create and manipulate a Majorana zero
mode localized on a defect is important for future applications in
quantum computations. Of course, before addressing such physics, the
pairing symmetry of candidate TSCs derived from TIs needs to be
elucidated, which is an important near-term challenge and requires
further advancements in materials synthesis techniques. 

On the theoretical front, classifications of possible topological phases
for various symmetries will continue to be important. In particular,
there are now intensive efforts on classifying and studying TCI phases
protected by various crystal symmetries \cite{95, 96, 97, 98, 99, 100, 101, 102,
103, 104, 105, 106, 107, 108, 118}, as well as
superconducting analogs of TCIs \cite{109, 110, 111, 112}. The role of crystal symmetry
in protecting topological nodal semimetals has also been studied
\cite{113, 114, 115, 116, 117}. The search for new TCI materials has attracted a great
interest. Theoretically predicted or proposed candidates include heavy
fermion compounds \cite{119,120}, transition metal oxides \cite{121}, and
antiperovskites \cite{122}. Regarding TSCs, 
mechanisms for odd-parity pairing in spin-orbit-coupled systems are being explored \cite{123a,124a, 125a, 126a, 127a},  
and their unusual topological properties are being studied \cite{129a, 130a, 131a, 132a, 135a,136a,138a,140a}. 
Importantly, the robustness of odd-parity superconductivity against disorder is found to 
be  parametrically enhanced by strong spin-orbit coupling \cite{141,142}. 
Last but not the least, a variety of new materials has recently  
been proposed as candidate time-reversal-invariant TSCs \cite{143, 144,145,148,149,150}, 
which makes this research field extremely active and lively.

Looking into the future, it remains to be seen whether strong
electron correlations can give rise to novel topological phases in
time-reversal-invariant systems, as was the case for fractional quantum
Hall effect in time-reversal-symmetry-broken systems. In both cases,
predictions of concrete candidates to realize newly conceived
topological phases are crucial for advancing the physics of topological
phases. To make the field of topological materials more interesting, it is
desirable that experimentalists discover unexpected topological phases
and phenomena in strongly correlated materials, and such serendipitous
discovery would lead to a major leap in our understanding.

\section{Acknowledgments}

Y.A. acknowledges Kouji Segawa, Alexey Taskin, Satoshi Sasaki, Zhi Ren,
Markus Kriener, Mario Novak, Fan Yang, Kazuma Eto, Takafumi Sato, Seigo
Souma, Takashi Takahashi, Yukio Tanaka, and Masatoshi Sato for
collaborations. L.F. thanks Arun Bansil, Erez Berg, Chen
Fang, Tim Hsieh, Charlie Kane, Hsin Lin, Junwei Liu, Ling Lu, Vidya
Madhavan, Karen Michaeli, Maksym Serbyn, Eveyln Tang and Ilijia
Zeljkovic for collaborations, and takes this opportunity to thank Charlie
Kane for introducing him to the field of topological insulators. Y.A.
was supported by JSPS (KAKENHI 25220708), MEXT (Innovative Area
``Topological Quantum Phenomena" KAKENHI), AFOSR (AOARD 124038). L.F.
is supported by DOE, Office of Basic Energy Science (DE-SC 0010526).

\begin{figure}
\begin{center}
\includegraphics[clip,width=13.5cm]{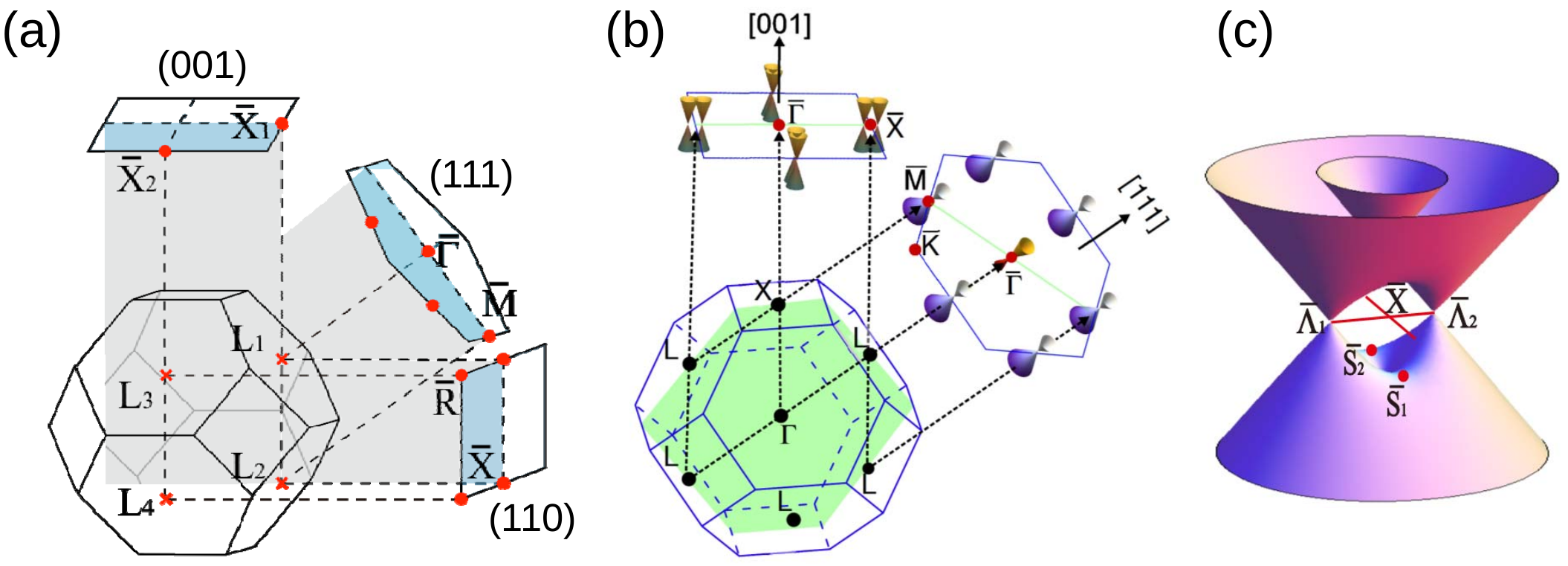}
\caption{Topological crystalline insulator (TCI). 
(a) High-symmetry points in the 3D Brillouin zone and in the projected
surface Brillouin zone for three different surfaces of the rock-salt
crystal structure. Adapted from Ref. \cite{Liu-Duan-Fu}; copyright (2013) 
by the American Physical Society.
(b) Locations of the Dirac cones in the (111) and (001) surface Brillouin zone. 
(c) Result of the tight-binding calculations for the dispersion of the
(001) double-Dirac-cone surface state. Adapted from Ref. \cite{Liu-Duan-Fu}; 
copyright (2013) by the American Physical Society.
}
\end{center}
\label{fig1}
\end{figure}

\begin{figure}
\begin{center}
\includegraphics[clip,width=11cm]{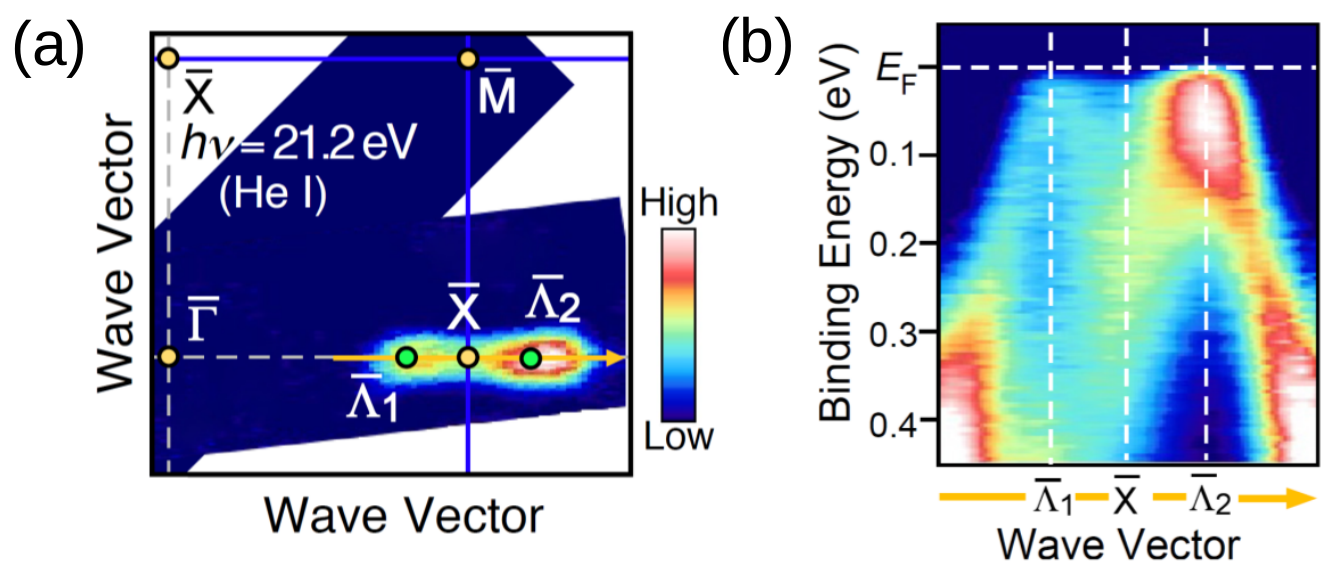}
\caption{Double-Dirac-cone surface state observed by 
angle-resolved photoemission spectroscopy (ARPES) experiments on
SnTe. 
(a) Distribution of the ARPES intensity at the Fermi energy $E_F$ in the
Brillouin zone ($k_x$ vs $k_y$). (b) The dispersion relations $E(k)$ when taken
as a slice through the Fermi surface found in panel (a); this slice is
taken along the yellow arrow indicated in panel (a). Adapted from
Ref. \cite{TanakaNP}.
}
\end{center}
\label{fig2}
\end{figure}

\begin{figure}
\begin{center}
\includegraphics[clip,width=13.5cm]{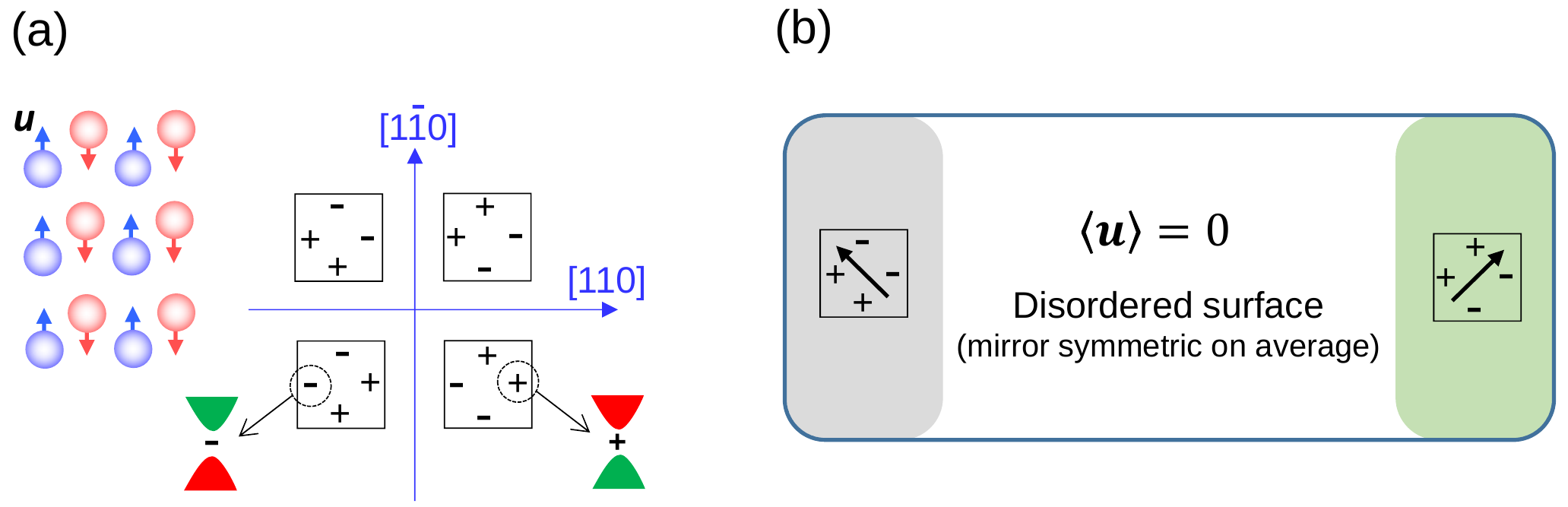}
\caption{
(a) Dirac mass generation due to ferroelectric structural distortion for
the SnTe-class of TCI materials. Both the magnitude and the sign of the
Dirac masses depend on the direction of the distortion ${\bf u}$, as depicted in
the figure. (b) Schematic picture to depict the robustness of the TCI
surface states against disorder; if disordered surface were localized,
there must be one helical mode localized on either left or right
boundary of the central disordered region, which would contradict mirror
symmetry.
}
\end{center}
\label{fig3}
\end{figure}

\begin{figure}
\begin{center}
\includegraphics[clip,width=11cm]{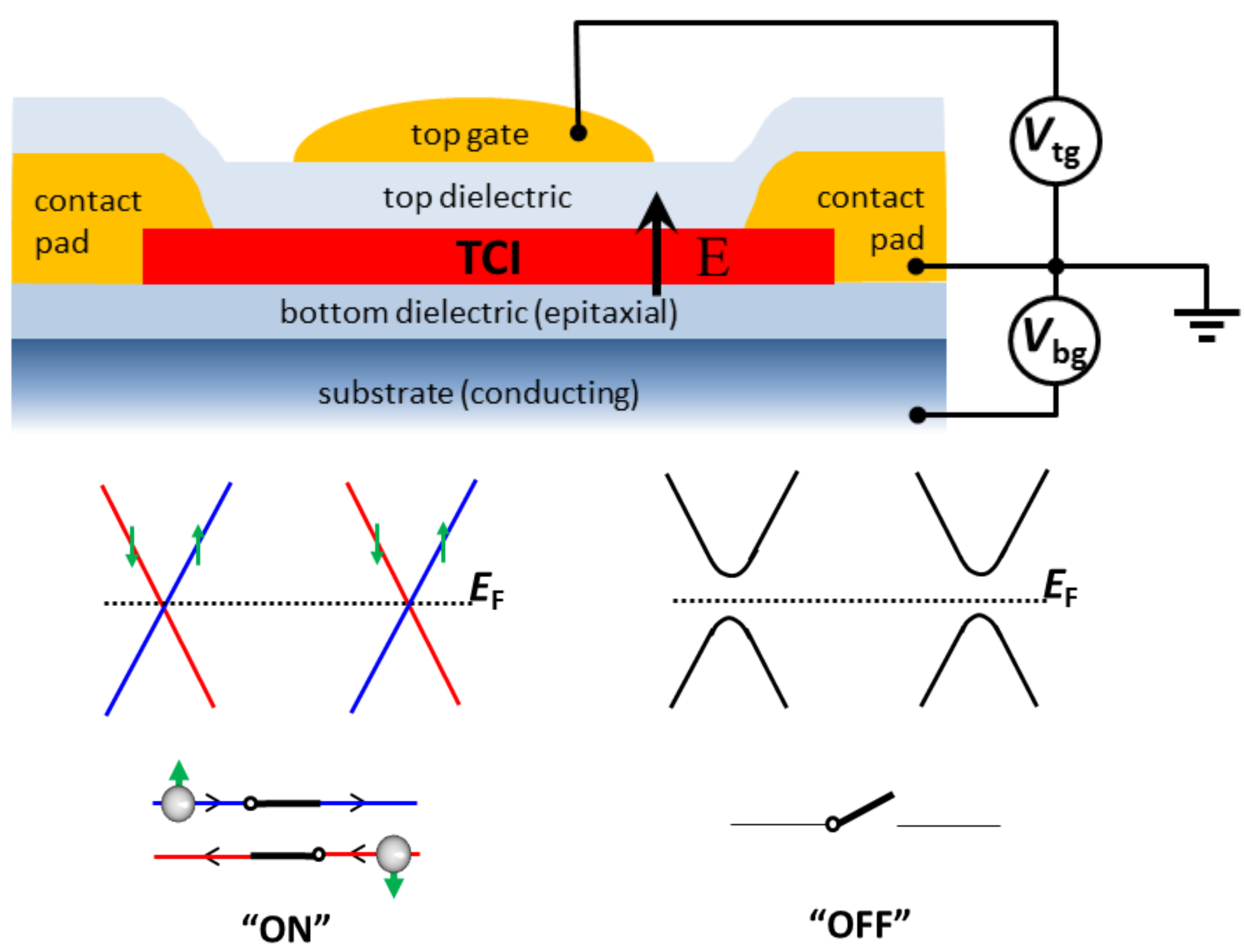}
\caption{
Possible TCI device to switch on and off the topological conduction
channel with electric field which breaks mirror symmetry with respect 
to the film's middle plane. Adapted from Ref. \cite{LiuDevice}.
}
\end{center}
\label{fig4}
\end{figure}

\begin{figure}
\begin{center}
\includegraphics[clip,width=13cm]{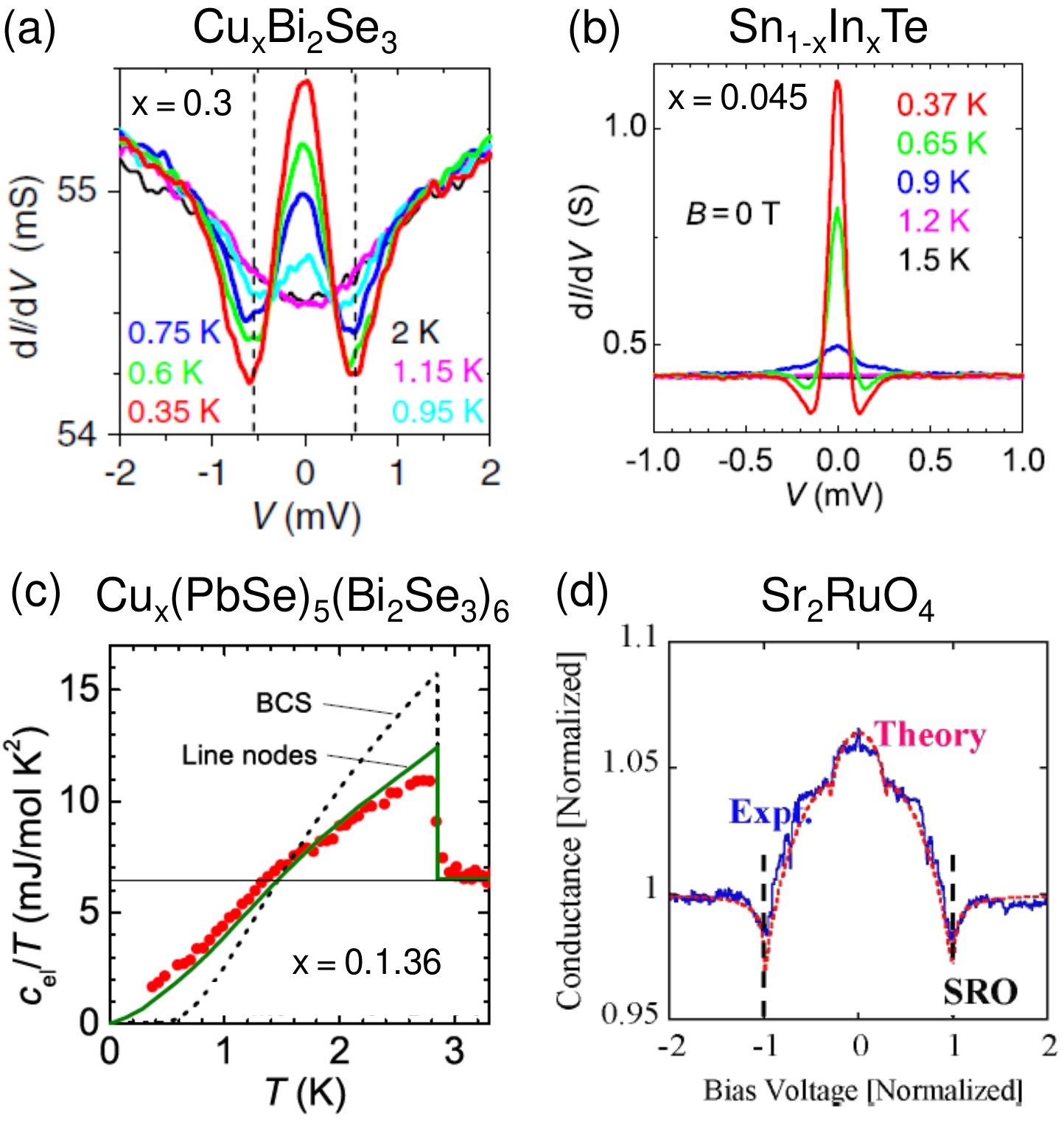}
\caption{Experiments on topological superconductor (TSC) candidates. 
(a) Zero-bias conductance peak observed in a point-contact 
spectroscopy of Cu$_x$Bi$_2$Se$_3$, which points to the existence
of surface Andreev bound states and makes this material a prime candidate
of the time-reversal-invariant TSC. Adapted from Ref. \cite{Sasaki2011}; 
copyright (2011) by the American Physical Society.
(b) Zero-bias conductance peak observed in the point-contact 
spectroscopy of Sn$_{1-x}$In$_x$Te, which is another candidate
of the time-reversal-invariant TSC. Adapted from Ref. \cite{Sasaki2012}; 
copyright (2012) by the American Physical Society.
(c) Temperature dependence of the electronic specific-heat of a
recently-discovered superconductor, Cu$_x$(PbSe)$_5$(Bi$_2$Se$_3$)$_6$
(CPSBS), which presents strong bulk signatures of unconventional
superconductivity with gap nodes. Adapted from Ref. \cite{Sasaki2014}; 
copyright (2014) by the American Physical Society.
(d) Zero-bias conductance peak observed in a tunneling spectroscopy of
Sr$_2$RuO$_4$, which is a prime candidate of the time-reversal-breaking
chiral TSC; calculated spectrum originating from the surface Andreev
bound state is also shown. Reprinted with permission from Ref. \cite{Kashiwaya}; 
copyright (2011) by the American Physical Society.
}
\end{center}
\label{fig5}
\end{figure}

\begin{table}[h]
\begin{tabular}{|l|c|c|c|c|}
\hline
Types of pairing in terms of field operators   &   Rep. & $P$ & $C_3$ & $M$ \\ 
  \hline   
  $\Delta_1: \; c^\dagger_{1\uparrow} c^\dagger_{1\downarrow} +   c^\dagger_{2\uparrow} c^\dagger_{2\downarrow}; \;  
  c^\dagger_{1\uparrow} c^\dagger_{2\downarrow} -  c^\dagger_{1\downarrow} c^\dagger_{2\uparrow} $ & $A_{1g}$ &  $+$  & $+$  & $+$ \\
  \hline
  $\Delta_2: \; c^\dagger_{1\uparrow} c^\dagger_{2\downarrow} +   c^\dagger_{1\downarrow} c^\dagger_{2\uparrow}$ & $A_{1u}$ & $-$ & $+$ & $-$ \\
  \hline
  $\Delta_3: \; c^\dagger_{1\uparrow} c^\dagger_{1\downarrow}  - c^\dagger_{2\uparrow} c^\dagger_{2\downarrow}$ & $A_{2u}$ & $-$ & $+$ & $+$ \\
  \hline
  $\Delta_4: \; (i c^\dagger_{1\uparrow} c^\dagger_{2\uparrow} -i   c^\dagger_{1\downarrow} c^\dagger_{2\downarrow}, 
   c^\dagger_{1\uparrow} c^\dagger_{2\uparrow} +   c^\dagger_{1\downarrow} c^\dagger_{2\downarrow})$ & $E_{u}$ & $(-, -)$ & $(x, y)$ & $ (+, -)$ \\
  \hline
\end{tabular}%
\caption{Four types of on-site pairing order parameters in the
two-orbital model for Cu$_x$Bi$_2$Se$_3$, which belong to the $A_{1g}$,
$A_{1u}$, $A_{2u}$ and $E_u$ irreducible representations of the D$_{3d}$
point group, as well as their transformation properties under point
group symmetry operations. Adapted from Ref. \cite{Fu-Berg}.
(Abbreviation: Rep., representation)}
\label{tab:SC}
\end{table}

\end{document}